\documentclass[seceqn,secthm]{elsart}

\usepackage{graphicx,natbib,amssymb,amsmath,lineno}

\usepackage{algorithm}
\usepackage{algorithmic}

\let\oldendproof\endproof
\def\endproof{\qed\oldendproof}

\renewcommand{\e}{\epsilon}

\newcommand{\I}{\mathcal{I}}
\renewcommand{\L}{\mathcal{L}}
\renewcommand{\S}{\mathcal{S}}
\newcommand{\vn}{\mathpzc{n}}
\DeclareMathAlphabet{\mathpzc}{OT1}{pzc}{m}{it}
\makeatletter
\def\elsartstyle{%
    \def\normalsize{\@setfontsize\normalsize\@xiipt{14.5}}
    \def\small{\@setfontsize\small\@xipt{13.6}}
    \let\footnotesize=\small
    \def\large{\@setfontsize\large\@xivpt{18}}
    \def\Large{\@setfontsize\Large\@xviipt{22}}
    \skip\@mpfootins = 18\p@ \@plus 2\p@
    \normalsize
} \@ifundefined{square}{}{} \makeatother

\begin{document}
\begin{frontmatter}

\title{Largest Empty Circle Centered on a Query Line}

\author[TRDDC]{John Augustine\corauthref{cor}},
\ead{john.augustine@tcs.com} \corauth[cor]{Corresponding author.}
\author[HMC]{Brian Putnam\thanksref{COLBY}}
\thanks[COLBY]{Work done as an undergraduate student at Colby
College, Waterville, Maine 04901, USA} \ead{BPutnam@hmccentral.com}
\author[TRDDC]{Sasanka Roy},
\ead{sasanka.roy@tcs.com}
\address[TRDDC]{Tata Research Development and Design Centre, Pune,
Maharashtra, India.}
\address[HMC]{HMC, 300 Chestnut St, Suite 101, Needham MA, 02492, USA}

%

%

\maketitle

\begin{abstract}
The Largest Empty Circle problem seeks the largest circle centered
within the convex hull of a set $P$ of $n$ points in $\mathbb{R}^2$
and devoid of points from $P$. In this paper, we introduce a query
version of this well-studied problem. In our query version, we are
required to preprocess $P$ so that when given a query line $Q$, we
can quickly compute the largest empty circle centered at some point
on $Q$ and within the convex hull of $P$.

We present solutions for two special cases and the general case; all
our queries run in $O(\log n)$ time. We restrict the query line to
be horizontal in the first special case, which we preprocess in $O(n
\alpha(n) \log n)$  time and space, where $\alpha(n)$ is the slow
growing inverse of the Ackermann's function. When the query line is
restricted to pass through a fixed point, the second special case,
our preprocessing takes $O(n \alpha(n)^{O(\alpha(n))} \log n)$ time
and space. We use insights from the two special cases to solve the
general version of the problem with preprocessing time and space in
$O(n^3 \log n)$ and $O(n^3)$ respectively.
\end{abstract}

\end{frontmatter}


\section{Introduction}

Facilities that pollute their surroundings are necessary evils. Our
cities and industrial towns need factories, dump grounds, dams, and
nuclear power plants. While we cannot eliminate them completely, we
would like to locate them far away from human dwellings. The same
problem arises on the flip side when locating, for instance, a
school far away from high crime areas and polluting facilities.
These scenarios have given rise to a well-studied class of problems
known as {\em Obnoxious Facility Location.} The most basic problem
in this class is the {\em Largest Empty Circle} problem, which takes
a set of points $P$ and asks for the largest circle with its center
inside the convex hull of $P$ and devoid of points in $P$.

In this paper, we study the placement of an obnoxious facility on a
region that can be modeled by a line. Consider the obnoxious
facility location that arises in disaster relief, in which planes on
a linear flight path must drop personnel in the disaster region.
They must, however, be dropped far away from points within the
region that pose imminent threat. We address this flavor of problems
by formulating a query version in which we are allowed to preprocess
the disaster region in a reasonable amount of time, and when queried
with a flight path, we can quickly provide the best place for
dropping relief personnel.

More formally, we study a query version of the Largest Empty Circle
problem in which we are given set of points $P=\{p_1, p_2, \cdots,
p_n\}$ strictly inside $[0,1]^2$. When given a query line $Q$, which
we sometime parameterize as $Q(t)$, we are to compute the largest
empty circle (abbreviated as LEC) with its center on $Q$ and within
the convex hull of $P$. The Largest Empty Circle Problem can have
multiple solutions, but conventionally requires us to report only
one solution. The query version we study  can also lead to multiple
solutions, so we keep the convention and limit our requirement to
{\em one} empty circle with the largest possible radius. For ease of
treatment, we simply call this {\em the} largest empty circle.

The Largest Empty Circle problem was first studied in the late 70s
and early eighties. Toussaint~\cite{Toussaint83} gave an $O(n \log
n)$ result and showed that it can be extended to the case where the
center is constrained to lie within a convex polygon. Bose and
Wang~\cite{BW02} have shown that $O(n \log n)$ time suffices even
when the center is to be constrained within a simple polygon. Chew
and Drysdale~\cite{CheDry86} studied the problem when the Voronoi
diagram and convex hull are give as part of the input. They showed
that  the LEC can be computed in $O(n)$ time. The query structure we
consider, i.e., requiring the center of the circle to be on a query
line, is not new either. For instance, Bose et al.~\cite{BLR08}
provide the {\em smallest enclosing circle} in $O(n \log n)$
preprocessing time, $O(n)$ space and $O(\log n)$ query time. For
more information on obnoxious facility location problems, the reader
is referred to the survey by Paola~\cite{cap99}.

In this paper, we present solutions for two special cases and the
general case. It is well known for the classical Largest Empty
Circle problem that the center of the LEC will either lie on a
Voronoi edge or the convex hull. In Lemma \ref{lem:vor}, we show
that this is true even when given a query line, implying that the
center is at an intersection of the query line with either a Voronoi
edge or the convex hull. A query line can intersect up to $O(n)$
Voronoi edges implying that trivial queries based on this insight
alone will not suffice in achieving sub-linear query times. The key
insight that is novel to this paper is the construction of a 3D
structure consisting of hyperbolic arcs such that when the query
line is dropped from $z=+\infty$, the point at which it lands on the
structure defines the center of the LEC. The three cases differ in
the way we modify the structure so that we can, in $O(\log n)$ time
for all cases, compute the landing point (and hence the center of
the LEC). This entails finding the upper envelope of a set of
functions for which we appeal to the concept of Davenport-Schinzel
sequences \cite{DavenportSchinzel65,Hershberger89,SharirAgarwal96}.

In the first special case, we restrict the query line to be
horizontal, i.e., of the form $y=y_c$, where $y_c \in [0,1]$ is a
constant. We preprocess it in $O(n \alpha(n) \log n)$ time and
space, where $\alpha(n)$ is the slow growing inverse of the
Ackermann's function. When the query line is restricted to pass
through a fixed point, the second special case, our preprocessing
takes $O(n \alpha(n)^{O( \alpha(n))}  \log n)$ time and space. In
both cases, we project the 3D structure consisting of hyperbolic
arcs to a 2D plane so that we can find the landing point.

For the general case, we assume without loss of generality that the
query line intersects the $x$-axis. In addition, we assume that for
any fixed query line, there are at most 3 LECs and consider more
than three occurrences to be degenerate. In Section \ref{sec:arb} we
describe how this degeneracy can be detected and removed. In Lemma
\ref{lem:SeqCard}, we show that the $x$-axis can be divided into
$O(n^3)$ intervals such that each is an instance of the ``query line
through a point" case. This easily leads to an $O(n^4\alpha(n)^{O(
\alpha(n))} \log n)$ time algorithm, but we use persistent data
structures \cite{DSST89} to reduce the preprocessing time to $O(n^3
\log n)$ with the data structure taking $O(n^3)$ space.

The paper is organized as follows. In Section \ref{sec:prelim}, we
provide some initial insights and a high-level framework for solving
all cases. In Section \ref{sec:hor}, we consider the special case in
which the query line is guaranteed to be horizontal. The second
special case in which the query line goes through a fixed point is
addressed in Section \ref{sec:origin}. Finally, we provide the
solution for the general case, i.e., arbitrary query lines, in
Section \ref{sec:arb}.

\section{Characteristics of the Solution}
\label{sec:prelim}

It has been known that the center of the LEC lies either on a
Voronoi edge or the convex hull of the points in $P$
\cite{Toussaint83}. This notion holds true in our situation also and
is captured by the following lemma. Note that since we require the
center of the LEC to lie within the convex hull, we only consider
portions of the Voronoi edges within the convex hull.

\begin{lem}
\label{lem:vor} The center of the LEC centered on the query line $Q$
will be the intersection point of $Q$ either with a Voronoi edge or
the convex hull.
\end{lem}

\begin{pf*}{Proof.} Let $q \in Q$ be the center of an LEC centered on $Q$. Let $p
\in P$ be a point closest to $q$. Consider the Voronoi cell
enclosing point $p$. The query line must pass through it (or at
least touch it at $q$) because otherwise, $q$ will be in some other
Voronoi cell and hence closer to some other point in $P$. Let
$Q(t^*)$ be the point on $Q$ closest to $p$. The distance from $p$
to $Q(t)$ will be a convex function with its minimum at $Q(t^*)$ and
strictly increasing on either direction of $t^*$. Therefore, the
point $q$ will be as far away from $Q(t^*)$ as it can go within the
Voronoi cell of $p$ and within the convex hull of $P$. This will
have to either be a Voronoi edge or the convex hull of $P$. \qed
\end{pf*}

\begin{figure}
\begin{center}
\includegraphics[width=90mm]{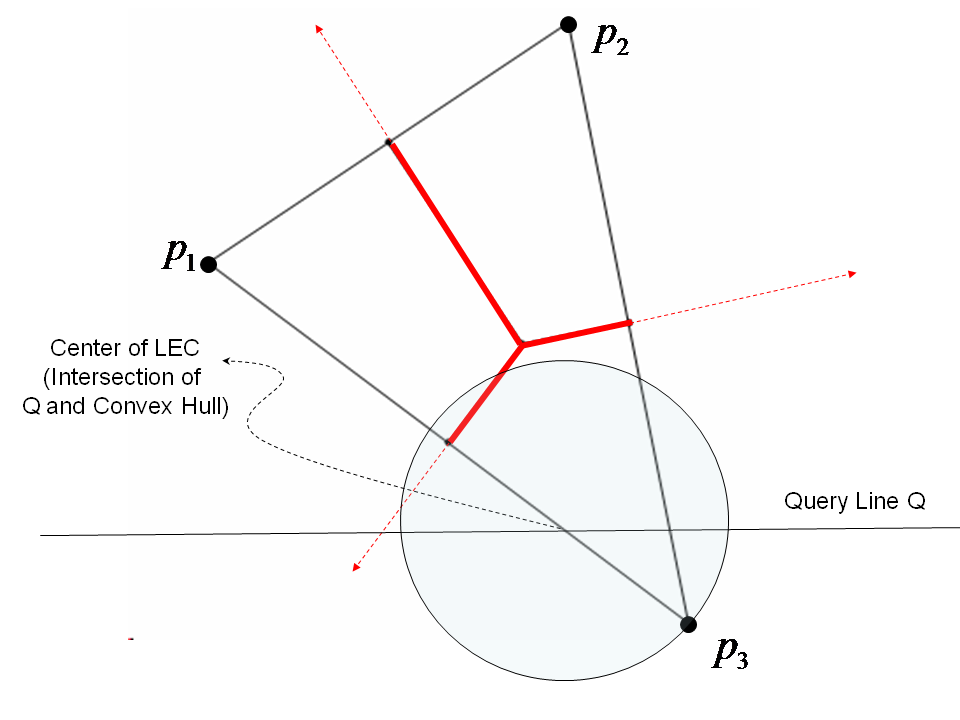}
\end{center}
\caption{The Query Line $Q$ missing all Voronoi edges inside the
convex hull. The intersections of $Q$ with the convex hull are
candidates for the center of the LEC.} \label{fig:ConvexHullCase}
\end{figure}

Lemma \ref{lem:vor} leads us to the straightforward algorithm of
checking all intersections of $Q$ with the Voronoi edges and the
convex hull edges. The convex hull intersections cannot be ignored
because in some cases, such as the instance shown in Figure
\ref{fig:ConvexHullCase}, the query line misses all the Voronoi
edges inside the convex hull. Note that the convex hull (and the
Voronoi diagram) can be constructed during preprocessing in $O(n
\log n)$ time.


\begin{figure}
\begin{center}
\includegraphics[width=90mm, bb=0 0 900 600]{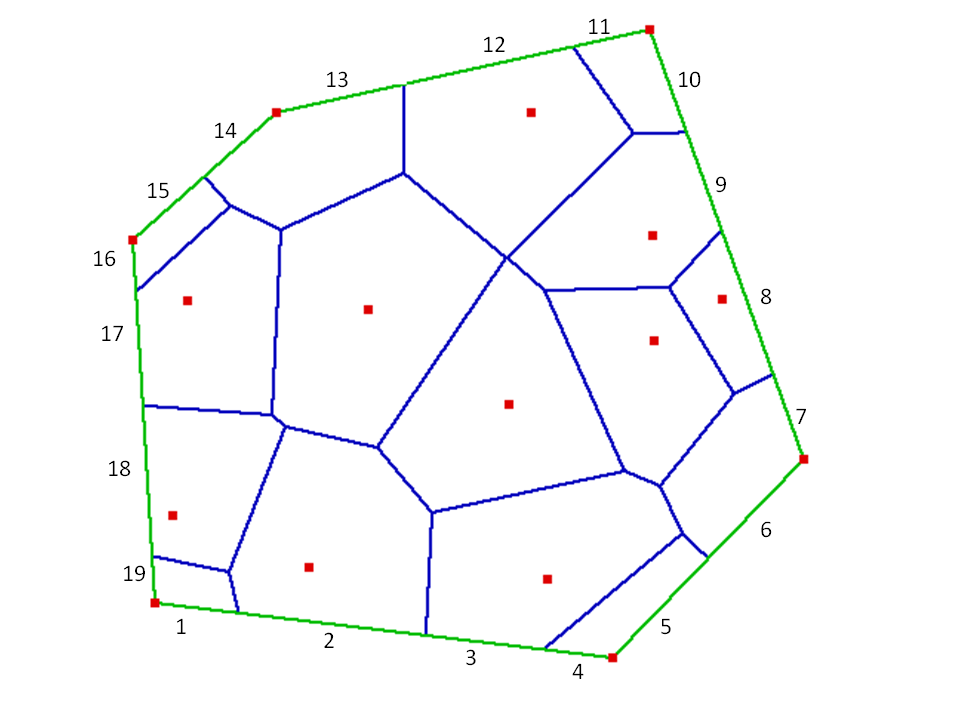}
\end{center}
\caption{The convex hull  and the internal segments of the Voronoi
edges of a set of points.} \label{fig:HullVoronoi}
\end{figure}

In the preprocessing step, as a consequence of Lemma \ref{lem:vor},
we consider the Voronoi edges and the convex hull as shown in Figure
\ref{fig:HullVoronoi}. More precisely, we don't consider the Voronoi
edges in full, but rather only those segments that are encompassed
by the convex hull. For the sake of convenience, we use the generic
term, Voronoi edges, to refer to these internal segments.
Additionally, we break the convex hull into segments delimited by
points in $P$ that are on the convex hull {\em and} the points where
the Voronoi edges intersect the convex hull. To illustrate, the
convex hull in Figure \ref{fig:HullVoronoi} is broken into 19
segments. Note that the number of such segments will be $O(n)$. For
the sake of uniformity, we treat all the $O(n)$ convex hull segments
and the $O(n)$ Voronoi edges as a single set $E$. Let $\vn \in O(n)$
be the cardinality of $E$.
 We construct a 3D
structure $H = \{ h_1, h_2, \cdots, h_{\vn}\}$, where each $h_j$ is
a hyperbolic arc corresponding to an $e_j \in E$. The hyperbolic
arcs are subsequently transformed in a manner that will allow us to
query in $O(\log n)$ time.  Algorithm \ref{alg:frame} gives us a
high-level framework  for preprocessing $P$. While the creation of
$H$ (line number \ref{lno:ConstructH} of Algorithm \ref{alg:frame})
is common to all versions of the problem that we study, the manner
in which the structure is transformed to a form that can be queried
(line number \ref{lno:TransformH} of Algorithm \ref{alg:frame}) is
quite different for each version and is progressively more
complicated. We explain the construction of $H$ in this section and
defer the description of the transformations to later Sections.
\begin{algorithm}
\caption{Framework for preprocessing $P$ for all cases.}
\floatname{algorithm}{Procedure} \label{alg:frame}
\begin{algorithmic}[1]

\STATE Construct $E$ consisting of the convex hull segments and the
internal Voronoi edges for $P$.

\STATE Construct the 3D structure $H$ consisting of hyperbolic arcs.
This is outlined subsequently in Section \ref{sec:prelim} and
illustrated in Figure \ref{fig:hyperbola}. \label{lno:ConstructH}

\STATE Process $H$ so that the landing point of the query line
dropped from $z=+\infty$ can be computed quickly.
\label{lno:TransformH}

\end{algorithmic}
\end{algorithm}

~

\begin{figure}
\begin{center}
\includegraphics[width=90mm, bb=0 0 900 600]{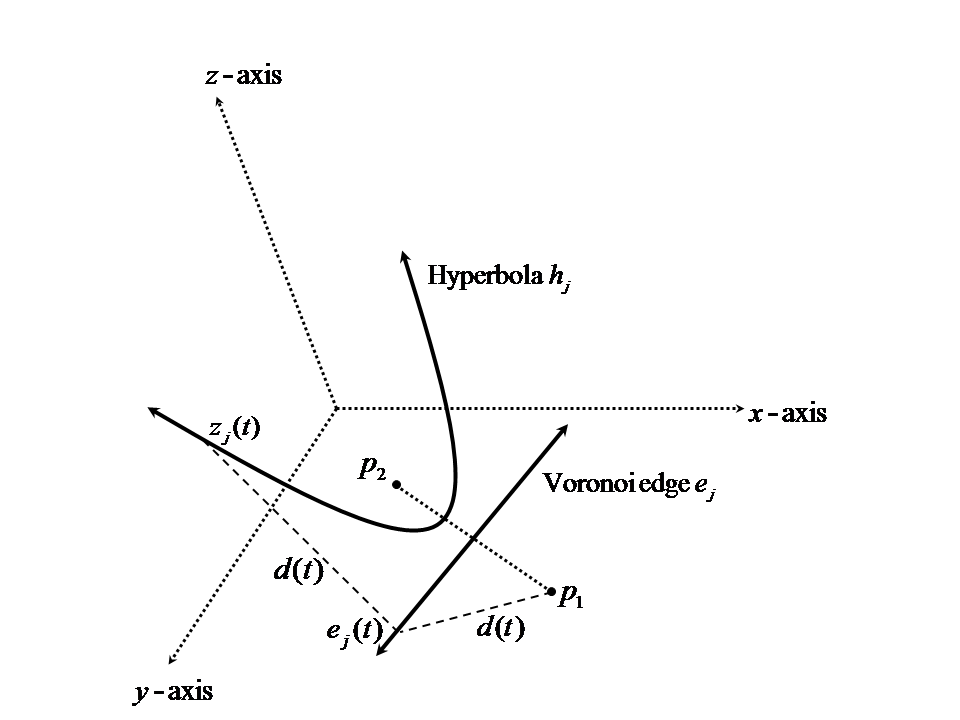}
\end{center}
\caption{Constructing the hyperbolic arc $h_j$ corresponding to a
single Voronoi edge $e_j$} \label{fig:hyperbola}
\end{figure}

For each Voronoi edge $e_j \in E$, we now describe how and why we
construct a corresponding hyperbolic arc $h_j$ directly above it as
shown in Figure \ref{fig:hyperbola}. Let $p_1$ and $p_2$ be the two
points that induce $e_j$. For convenience, we consider $e_j(t)$ to
be the parameterized representation of the point $(x_t,y_t)$ on
$e_j$ that is $t$ units from the intersection of the line segment
$p_1p_2$ and the (possibly extended) Voronoi edge $e_j$.
Intuitively, each point $(x_t,y_t,z_t)$ of the hyperbolic arc $h_j$
is the elevation of the point $(x_t,y_t) \in e_j$ in the $+z$
direction to a height $z_t$ that equals the euclidean 2D distance
from $(x_t,y_t)$ to either $p_1$ or $p_2$. Let $d(t)$ be the
euclidean distance from $p_1$ (or equivalently $p_2$) to $e_j(t)$.
The height of  the hyperbolic arc $z_t$ corresponding to the point
$e_j(t)$ on $e_j$ is equal to the distance $d(t)$. Therefore, $$z_t
= d_j(t) = \sqrt{t^2 + \left (\frac{dist(p_1,p_2)}{2}\right )^2},$$
where $dist(p_1,p_2)$ is the euclidean distance between $p_1$ and
$p_2$. Hence the arc is hyperbolic.

We note that the above description also holds for the convex hull
segments in $E$. Each convex hull segment is closest to at most one
point because of the way we have segmented the convex hull. This
closest point induces the hyperbolic arc the same way the two points
induce the arc for Voronoi edges. Note that this hyperbolic arc will
be a straight line for those convex hull segments that are incident
on a point $p$ in the convex hull of $P$. This is a natural
consequence of the way we construct the hyperbolic arcs and does not
pose a problem to our algorithms.

While Figure \ref{fig:hyperbola} shows the construction of the
hyperbolic arc
 on an infinitely long Voronoi
edge, our problem is restricted to finding the LEC with center
within the convex hull, we will only consider the (finite) edges in
$E$. Such finite edges will only induce hyperbolic arcs rather than
the full hyperbolas. Note also that each hyperbolic arc can be
constructed in $O(1)$ time if the points that induce the
edge\footnote{Of course, the hyperbolic arcs above the convex hull
segment are only induced by one point.} and the extents of the edge
(i.e., the $t$ values between which the hyperbolic arc is defined)
are given to us.

One way to interpret this structure in light of Lemma \ref{lem:vor}
is to drop  our query line $Q$ from $z=+\infty$ onto the hyperbolic
arc structure and report the center for the LEC corresponding to the
point at which it touches some hyperbolic arc and projected straight
down onto the $z=0$ plane.

\section{Horizontal Query Line}
\label{sec:hor} In this section, we assume that the query line $Q$
will be of the form $y=y_c$, where $y_c$ is some constant. recall
that the points in $P$ lie strictly in $[0,1]^2$. The preprocessing
is outlined in Algorithm \ref{alg:PreHor}.

\begin{algorithm}
\caption{Preprocessing $P$ for horizontal query lines.}
\floatname{algorithm}{Procedure} \label{alg:PreHor}
\begin{algorithmic}[1]

\STATE Construct $E$ consisting of the convex hull segments and the
internal Voronoi edges for $P$.

\STATE Construct the 3D structure $H$ consisting of hyperbolic arcs
as outlined in Section \ref{sec:prelim}

\STATE Project each hyperbolic arc in $H$ orthographically onto the
$x=0$ plane. \label{lno:proj}

\STATE Find the upper envelope of the projected hyperbolic arcs
using the algorithm outlined in \cite{SharirAgarwal96}.
\end{algorithmic}
\end{algorithm}

In step number \ref{lno:proj} of Algorithm \ref{alg:PreHor}, each
point $(x,y,z)$ in a hyperbolic arc $h_j$ will be projected
(orthographically) onto the point $(0, y, z)$. The orthographic
projection preserves the hyperbolic nature of the arcs. Therefore,
each hyperbolic arc $h_j$ becomes a hyperbolic arc $h'_j$ on the
$x=0$ plane. Let $H' = \{ h'_1, h'_2, \cdots, h'_{\vn}\}$. Recall
that we are only concerned with the first point at which the query
line, $y=y_c$, touches the structure when dropped from $z=+\infty$
onto $H$. It is easy to see that this corresponds to the upper
envelope curve in $H'$ at $y=y_c$. Hence, we need the upper envelope
of $H'$. Any two projected hyperbolas in $H'$ can intersect at a
maximum of two points and are partially defined owing to the fact
that all hyperbolic arcs in $H$ are constructed above the $[0,1]^2$
region. Therefore, we appeal to the definition of Davenport-Schinzel
sequences and Theorem \ref{thm:ds} stated by Sharir and Agarwal
\cite{SharirAgarwal96} and restated here in an abridged manner to
capture our requirement. (Please refer to the excellent exposition
by Sharir and Agarwal \cite{SharirAgarwal96} for more information
about Davenport-Schinzel sequences and their application in finding
lower and upper envelopes.)

\begin{defn} \cite{SharirAgarwal96}
Let $A$ be an alphabet with $\vn$ characters and $s>0$ be an integer
constant. A sequence $U = {a_1, a_2, \cdots, a_m}$, where each $a_i
\in A$, is an $(\vn, s)$ {\em Davenport-Schinzel sequence} if it
satisfies the following conditions:
\begin{enumerate}
  \item $a_i \ne a_{i+1}$ for each $i < m$, and
  \item there do not exist $s+2$ indices $i_1, i_2, \ldots,
  i_{s+2}$, where $i_1 < i_2 < \cdots < i_{s+2}$,
  such that
  $$ a_{i_1} = a_{i_3} = a_{i_5} = \cdots  =a, \quad a_{i_2} = a_{i_4} = a_{i_6} = \cdots = b $$
  and $a \ne b$.
\end{enumerate}
\end{defn}

\begin{defn} \cite{SharirAgarwal96}
$\lambda_s(\vn) = \max_U {|U|}$, where $U$ is an $(\vn, s)$
Davenport-Schinzel sequence.
\end{defn}

\begin{thm} \cite{SharirAgarwal96,Hershberger89}
\label{thm:ds} Given a set $H'$ of $\vn$ partially defined
univariate functions,  the sequence of functions forming the upper
envelope is a $(\vn,s+2)$ Davenport Schinzel sequence, where $s$ is
the number of points at which two functions $h'_1\in H'$ and
$h'_2\in H'$ can meet. It can be computed in $O(\lambda_{s+1}(\vn)
\log \vn)$ time.
\end{thm}
Since our hyperbolas meet at 2  points at most, $s=2$. Further, the
hyperbolas are partially defined. Therefore, from Theorem
\ref{thm:ds} and the upper bounds on $\lambda$ functions given in
\cite{SharirAgarwal96}, we know that $\lambda_{2+1} (\vn)=\lambda_3
(\vn) = O(n \alpha(n))$, where $\alpha(n)$ is the slow-growing
functional inverse of the Ackermann's function. This directly leads
us to the following corollary.

\begin{cor}
The upper-envelope of the set of partially defined hyperbolas $H'$,
can be computed in $\lambda_3(n) \log n = O(n \alpha(n) \log n)$.
\end{cor}
\begin{pf*}{Proof.}
In $H'$, all the hyperbolas are oriented upward and hence intersect
at most at 2 points. Further, since they are partially defined, the
proof follows from Theorem \ref{thm:ds}. \qed \end{pf*}

The upper envelope of $H'$ will be a sequence  of maximal intervals
of $y$-values such that within each interval we have a single
function from $H'$ that dominates. Hence, when we get a query line
of the form $y=y_c$, we merely search for the interval  that
contains $y_c$. This takes $O(\log n)$ time. Let $e_j\in E$ be the
edge that induced the hyperbola corresponding to the resulting
interval. The intersection of $Q$ and $e_j$ is the required center
of the LEC.

\begin{thm}
\label{thm:hor} Given a set $P$ with points in $[0,1]^2$, we can
preprocess $P$ in $O(n \alpha(n) \log n)$ time and space such that
when we are given a query line $Q$ of the form $y=y_c$, where
$y_c\in[0,1]$, we can report the LEC centered on $Q$ in $O(\log n)$
time.
\end{thm}

\section{Query Line Through a Fixed Point}
\label{sec:origin} In this section, we study the special case where
the query line $Q$ passes through a pre-specified point called the
pivot. Without loss of generality, we assume that the origin is the
pivot. We work out the essential details first assuming that $P\in
[0,1]^2$; this restricts the pivot to the bottom left corner.
Subsequently, we will show how this can be generalized to $P\in
[-1,1]^2$.

In this section, we assume that the set of edges $E$ and the
corresponding set of hyperbolic arcs $H$ are already constructed. We
say that a query line $Q$ {\em lands} on $e_j\in E$ at angle
$\theta$ if it makes an angle $\theta$ with the $x$-axis and an LEC
is centered at the intersection of $Q$ and $e_j$. In other words,
when $Q$ is dropped from $z=+\infty$, the hyperbolic arc $h_j$
corresponding to $e_j$ is the first hyperbolic arc it lands on.
Drawing from this imagery of the query line landing on hyperbolic
arcs, we use the phrases ``landing on the edge $e_j$" and ``landing
on the hyperbolic arc $h_j$" interchangeably. In case it lands on
more than one hyperbolic arc, we simply consider one of them.

Consider an interval $[\theta_1, \theta_2]$ such that for all
$\theta_1 \leq \theta \leq \theta_2$, $Q$ lands on a hyperbolic arc
$h_j$ (or equivalently on the corresponding Voronoi edge $e_j$).
Extending our previous definition, we say that $Q$ lands on $h_j$
(or $e_j$) in the angular interval $[\theta_1, \theta_2]$.

\begin{figure}
\begin{center}
    \includegraphics[width=60mm]{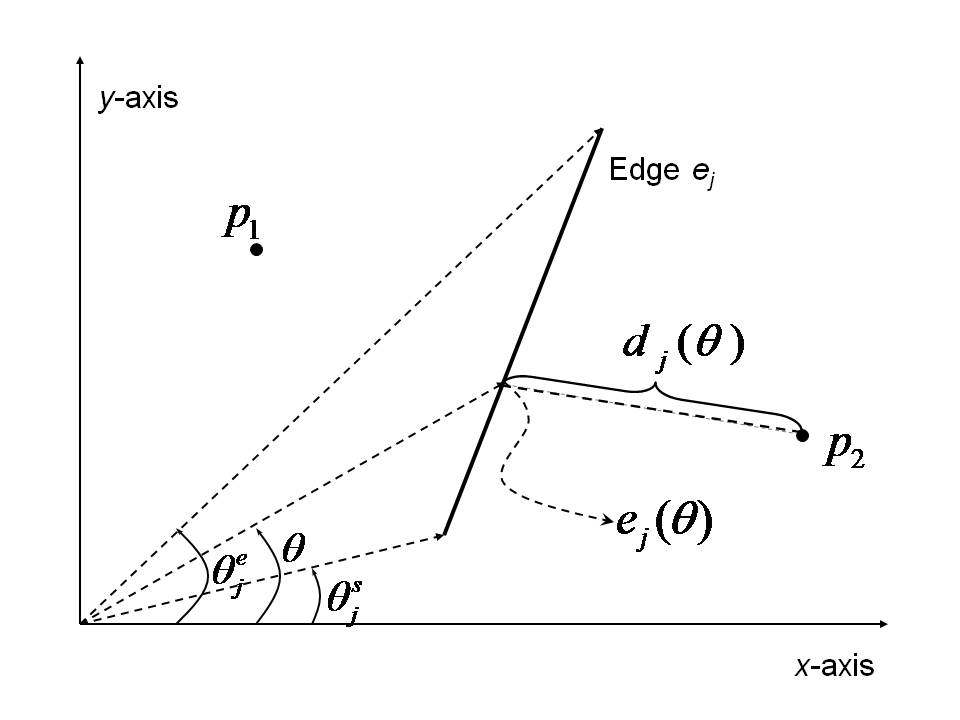}
    \includegraphics[width=60mm, bb=0 0 900 600]{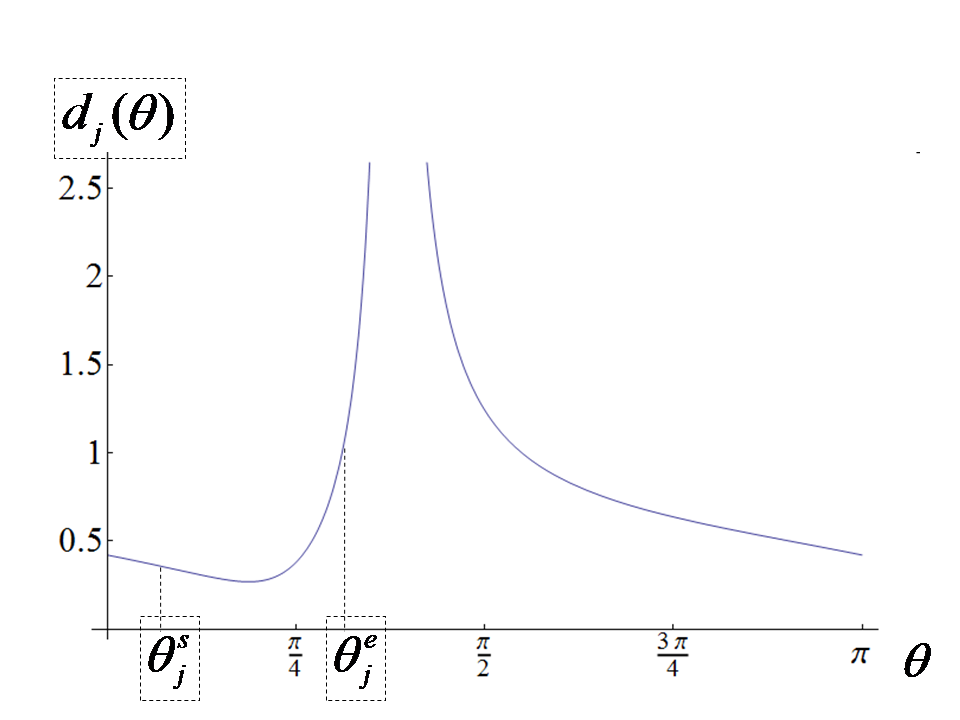}
\end{center}
\caption{Computing the $d_j(\theta)$ function for a single edge $e
\in E$. Note that the $d_j(\theta)$ function shown on the right
roughly corresponds to the edge shown on the left. The portion of
the curve we need is limited to the angular interval $[\theta_j^s,
\theta_j^e]$.} \label{fig:AngularProj}
\end{figure}

To aid in computing the edge on which a query line lands, we first
compute a function $d_j(\theta)$ for each edge $e_j$ taken
individually. We omit the subscript when it is clear from the
context. Intuitively, it is the height at which a query line $Q$
making an angle $\theta$ with the $x$-axis touches the hyperbolic
arc $h_j$ constructed over $e_j$.
More precisely, if $[\theta_j^s,\theta_j^s]$ is the maximal range of
angle such that a query line making an angle $\theta \in
[\theta_j^s,\theta_j^s]$ intersects edge $e_j$ at a point denoted
$e_j(\theta)$, then
\[
d_j(\theta) =
\begin{cases}
dist(e_j(\theta), p_1), &\text{if $\theta_j^s \le \theta \le \theta_j^e$;}\\
0, &\text{Otherwise.}
\end{cases}
\]
In the above equation, $dist(e_j(\theta), p_1)$ is used to denote
the 2D euclidean distance between $e_j(\theta)$ and $p_1$.The upper
envelope of all these functions, $D(\theta) = \sup_j
d_j(\theta)$\footnote{We use $\sup(X)$ to denote the supremum over a
set $X$ of real values, which is defined to be the smallest real
value that is greater than or equal to every $x \in X$.}, provides
us the radius of the LEC given a query line $Q$ making an angle
$\theta$ with the $x$-axis.
%
With some algebraic and geometric manipulations, we get
$d_j(\theta)$ to be of the form
\begin{equation} d_j(\theta) = \sqrt{\frac{a_1 \theta^2
+ a_2 \theta +a_3}{(\theta - a_4)^2}}, \label{eqn:Dfunction}
\end{equation}
where $a_1$, $a_2$, $a_3$, and $a_4$ are constants dependent on the
point in $P$ that induces the edge $e_j$.

When given a query line $Q$ passing through the origin and making an
angle $\theta$, we intuitively ``drop" the query line from
$z=+\infty$ and consider the first hyperbolic arc it touches. This
will be the hyperbolic arc with the largest $d(\theta)$ value. Since
the query line can have an arbitrary angle $\theta$, we compute the
upper envelope $D(\theta)$ of the hyperbolic arcs in the
preprocessing step. In the query phase, we can search $D(\theta)$ to
find the appropriate hyperbolic arc that $Q$ lands on. Finally, we
define the {\em landing sequence} of hyperbolic arcs, denoted $\L$,
to be $(l_1, l_2, \cdots, l_k)$, to be the sequence of hyperbolic
arcs that are encountered as a query line is swept from $0$ to
$\pi/2$.



\begin{lem}
\label{lem:SeqCard} Given a set $P$ of $n$ points, each point lying
in $[0,1]^2$, and a query line $Q$ that is restricted to pass
through the origin, the landing sequence of $Q$ is an $(O(n), 6)$
Davenport-Schinzel sequence consisting of partially defined
functions.
\end{lem}
\begin{pf*}{Proof.}
The upper envelope curve $D(\theta)$ that is constructed from the
individual $d_j(\theta)$ curves of the edges in $E$ determines the
landing sequence. (Note that $|E| = \vn\in O(n)$). The complexity of
$D(\theta)$ in turn depends on the  possible number of intersections
between any two $d(\theta)$ curves. If we equate the two $d(\theta)$
functions (given in Equation \ref{eqn:Dfunction}) and solve the
resulting fourth degree equation for $\theta$, we will get up to 4
roots. Therefore, theoretically we can have at most four angles at
which any two $d(\theta)$ functions can intersect. Also, we know
that our hyperbolic arcs (and the useful range of the $d(\theta)$
functions) are of limited size. Therefore, the landing sequence will
be an $(\vn, 6)$ Davenport-Schinzel sequence consisting of partially
defined functions. \qed \end{pf*}

Therefore, from \cite{SharirAgarwal96}, we can construct the upper
envelope of the $d(\theta)$ functions for all the hyperbolic arcs
(and hence the  landing sequence) in $O(\lambda_{5}(\vn) \log \vn)=
O(\vn \alpha(\vn)^{\alpha(\vn)} c^{\alpha(\vn)} \log \vn)$ time,
where $c>0$ is some constant. Since $\vn \in O(n)$ and
$c^{\alpha(\vn)}$ grows slower than $\alpha(\vn)^{\alpha(\vn)}$ in
the asymptotic sense, we can rewrite the running time as $O(n
\alpha(n)^{O(\alpha(n))} \log n)$ and up to an equal amount of
space. Although Lemma \ref{lem:SeqCard} indicates that any two
$d(\theta)$ functions intersect at up to 4 points, we have been
unable to realize this in an example. We believe that the sequence
will in reality be simpler, but we don't have a proof for it. The
preprocessing steps are outlined in Algorithm \ref{alg:PreOrigin}.

\begin{algorithm}
\caption{Preprocessing $P$ for query lines through the origin.}
\floatname{algorithm}{Procedure} \label{alg:PreOrigin}
\begin{algorithmic}[1]

\STATE Construct the Convex Hull and the Voronoi diagram for $P$.

\STATE Construct hyperbolic arcs, one for each Voronoi edge and for
each convex hull segment.

\STATE Compute the $d_{\theta}$ function for each hyperbolic arc
according to Equation \ref{eqn:Dfunction}.

\STATE Compute the upper envelope $D(\theta)$ of the set of all
$d_{\theta}$ functions using the algorithm outlined in
\cite{SharirAgarwal96}.\label{lno:UpperEnvelope} Additionally, store
the angles at which the transitions occur in $D(\theta)$.
\end{algorithmic}
\end{algorithm}

Given the upper envelope of the $d(\theta)$ curves and the angles at
which the transitions occur in the upper envelope, we can, in
$O(\log n)$ time, find the exact hyperbolic arc on which a given
query line (passing through the origin and making an angle $\theta$
with the $x$-axis) lands. Substituting the angle $\theta$ in the
$d(\theta)$ function for the hyperbolic arc, we can get the radius
of the largest enclosing circle. The intersection of the query line
$Q$ with the edge it lands on is the center of the LEC.

\subsection{Pivot in Arbitrary Location}
So far, we have worked
under the assumption that the pivot point, i.e., the point through
which the query line must pass, is the origin and the points are in
$[0,1]^2$. However, this does not ensure the generality of the
solution. In particular, what if the pivot point needs to be inside
the convex hull of $P$? We can address this by placing the points in
$P$ in $[-1,1]^2$. Now, without loss of generality, the pivot can
continue to be the origin. We again need to compute the $d(\theta)$
function for each edge in $E$, but this time, we sweep a ray
(starting at the origin) about the origin for the entire $2\pi$
radians. This will not affect the asymptotic running time adversely
because any two $d(\theta)$ functions will again have at most 4
intersections. A consequence of using rays instead of lines for
computing the $d(\theta)$ functions is that when the query line
makes an angle $\theta$ with the $x$-axis, we have to check the
upper envelope at $\theta$ and $\pi + \theta$.

\section{Arbitrary Query Line}
\label{sec:arb} In this section, we consider the general version of
the problem in which the query line can be arbitrary. Recall that we
assume that the points are strictly in $[0,1]^2$. Any query line
that intersects the convex hull of $P$ must also intersect two edges
of the $[0,1]^2$ square. We assume that our query line $Q$
intersects the edge on the $x$-axis; we don't compromise on
generality because we can repeat this preprocessing for  other edges
without any asymptotic increase in time or space.

Our approach for this general version builds on the special case
studied in Section \ref{sec:origin} where $Q$ passes through the
origin. Like before, we construct a set $H$ of  hyperbolic arcs in
3D space and seek the point at which $Q$ lands.  We show that the
$x$-axis (between $[0,1]$) can be partitioned into $O(n^3)$ maximal
segments such that query lines intersecting a given segment induce
the same landing sequence. Note that the upper-envelopes will vary
depending on the exact point at which $Q$ intersects the $x$-axis,
but we seek maximal segments in which the landing sequence will
remain unchanged.

The algorithm described in Section \ref{sec:origin} will not suffice
because it requires the upper envelope, which lends itself to binary
searching. The landing sequence is merely a sequence of hyperbolic
arcs. The angular range in which each arc dominates is not included.
So it is not possible to determine whether the query line intersects
the upper envelope before or after the middle element in any portion
of the landing sequence. Therefore we start with a detour to show
how the algorithm described in Section \ref{sec:origin} can be
modified to work with the landing sequence and without the full
upper envelope. Subsequently, we show that the landing sequences
changes incrementally as the point at which the query line
intersects the $x$-axis moves along the $x$-axis. This allows us to
store the incrementally changing landing sequences in a persistent
data structure, which can be queried in $O(\log n)$ time.

\subsection{Querying with the landing sequence}
\label{sec:QuerySequence} In this section, we show that the
algorithm described in Section \ref{sec:origin} can be modified to
work even when restricted to using the landing sequence $\L$ alone
and not the upper envelope $D(\theta)$. For binary search to work,
we need to ask whether the query line at angle $\theta$ intersects
the upper envelope before or after the middle element of the
sequence and recurse either left or right according to the response.
As mentioned earlier, we cannot answer this question because we
don't know the angle at which the $d(\theta)$ function of each
hyperbolic arc in the landing sequence starts to dominate over the
previous element in the landing sequence. We overcome this
limitation by storing a little more information that neither
increases the running time nor the space (in the asymptotic sense).

From the definition, in an $(\vn,6)$ Davenport-Schinzel sequence
such as the landing sequence, any two elements $a$ and $b$ can occur
in order  at most three times. Therefore the $\cdots, a,b,\cdots$
consecutive pair can also occur at most 3 times. We call them the
meetings of $a$ and $b$.  Given a pair $a$ and $b$ and a point on
the $x$-axis through which our query line passes, we can find the
(at most) three meeting angles at which $a$ hands over to $b$ in
$O(1)$ time by simply equating the two $d(\theta)$ functions of $a$
and $b$ and solving for $\theta$. Of the four roots and taking the
extremities of the partial functions, at most three will correspond
to $a$ handing over to $b$. Therefore, in the preprocessing phase,
we store the meeting number (i.e., either first, second or third)
along with each meeting without any increase in the asymptotic
running time or space. This allows us to treat the three meetings
independently. In Algorithm \ref{alg:PreOrigin}, we add the
following line after Line number \ref{lno:UpperEnvelope}.
``Construct the landing sequence $\L=(l_1, l_2, \cdots, l_|\L|)$
from the upper envelope and additionally store the meeting numbers
as a separate sequence $M=(m_1, m_2, \cdots, m_{|\L|-1})$, where
$m_j$ is the meeting number of the pair $l_j, l_{j+1}$."

In the query phase, we perform a binary search on $\L$ (in
conjunction with $M$) by asking the following question: is the angle
$\theta$ that the query line makes with the $x$-axis to the left or
right of the ${m_j}$th meeting of $l_j l_{j+1}$, where the $m_j$th
meeting of $l_j l_{j+1}$ is the central pair in the portion of the
landing sequence considered in the current recursion. This can be
answered in $O(1)$ time, thereby establishing an $O(\log n)$ query
time.

\subsection{Partitioning the $x$-axis}
Let $\L_c$ be the landing sequence for query lines passing through
$(c,0)$. Along the $x$-axis, we encounter several maximal intervals
of the form $[x_1,x_2]$, $0\leq x_1 \leq x_2 \leq 1$, during which
the landing sequence $\L_c$ remains unchanged for all $c \in
(x_1,x_2)$. As we walk along the $x$-axis from $x=0$ to $x=1$, the
landing sequence changes at a finite number of points that we call
{\em events}. We now ask three questions:
\begin{enumerate}
    \item what are the possible types of changes that an event can
    induce in the landing sequence,
  \item how many such events exist in $[0,1]$, and
  \item how do we compute the events?
\end{enumerate}
We address these questions in a series of lemmas. In order to prove
these lemmas, we consider the following situation to be degenerate:
four or more LECs occur on a single query line.  In other words, the
degenerate case happens when a ray lands simultaneously on  four or
more hyperbolic arcs. Subsequently, we will briefly show how to
detect and accommodate this degeneracy by slightly perturbing the
points in $P$.

We note that similar situations have been studied previously along
with observations that are similar to ours. Bern et al.
\cite{BerDobEpp94}, for instance, study the changes in the topology
of a 3D scene when viewed along a straight line flight path. Suppose
we walk along the $x$-axis starting from the origin toward $x=1$ in
order to track the changes that we encounter in the landing
sequence. Let $x=c$ be an event point. For some small $\e>0$,
$\L_{c-\e} \ne \L_{c+\e}$. Let us first assume that the change is an
insertion of hyperbolic arcs at a single location in the sequence.
We will see that either one or two arcs are inserted. Note that an
insertion when walking in one direction is a deletion when walking
in the opposite direction, hence our assumption does not affect
deletions.
Suppose the sequence of hyperbolic arcs inserted at $x=c$ is $\I$.
We can break up $\L_{c-\e}$ and $\L_{c+\e}$ into subparts consisting
of hyperbolic arc sequences (denoted by $\S_1$ and $\S_2$), and
single hyperbolic arcs (denoted by $h$, $h_1$, and $h_2$) in two
possible ways as shown below.\\
\noindent {\bf Case A:} $\L_{c-\e} = \S_1+h+\S_2$ and $\L_{c+\e} = \S_1+h+\I+h+\S_2$, or \\
\noindent {\bf Case B:} $\L_{c-\e} = \S_1+h_1+h_2+\S_2$ and $\L_{c+\e} = \S_1+h_1+\I+h_2+\S_2$, \\
where ``$+$" is the concatenation operator. In other words,  $\I$ is
either inserted in the middle of an existing hyperbolic arc (Case A)
or at the meeting point of two hyperbolic arcs (Case B). Lemma
\ref{lem:possibilities} provides us the answer to the first
question: what types of changes can we encounter?

\begin{lem}
If we further divide the cases defined above as shown in the table
below, Case A* and Case B* cannot occur for non-degenerate input.

\begin{tabular}{|c|c|c|}
  \hline
  \textbf{Case name} & \textbf{Description} & \textbf{Can occur?} \\ \hline \hline
  Case A1 & Case A and $|\I|=1$ & Yes (See Figure \ref{fig:CaseA1}) \\ \hline
  Case A2 & Case A and $|\I|=2$ & Yes (See Figure \ref{fig:CaseA2})\\  \hline
  Case A* & Case A and $|\I|>2$ & No \\  \hline
  Case B1 & Case B and $|\I|=1$ & Yes (See Figure \ref{fig:CaseB1})\\  \hline
  Case B* & Case B and $|\I|\geq2$ & No \\  \hline
  \hline
\end{tabular}
\label{lem:possibilities}
\end{lem}
\begin{pf*}{Proof.} Cases A1, A2, and B1 are shown schematically in Figures
\ref{fig:CaseA1}, \ref{fig:CaseA2}, and \ref{fig:CaseB1}
respectively. Note that each figure highlights a small portion of
the the hyperbolic arcs viewed orthographically as explained in
Section \ref{sec:origin}. We prove the impossibility of occurrence
of the two cases separately.

\begin{figure}
\begin{center}
    \includegraphics[width=50mm, bb=0 0 900 600]{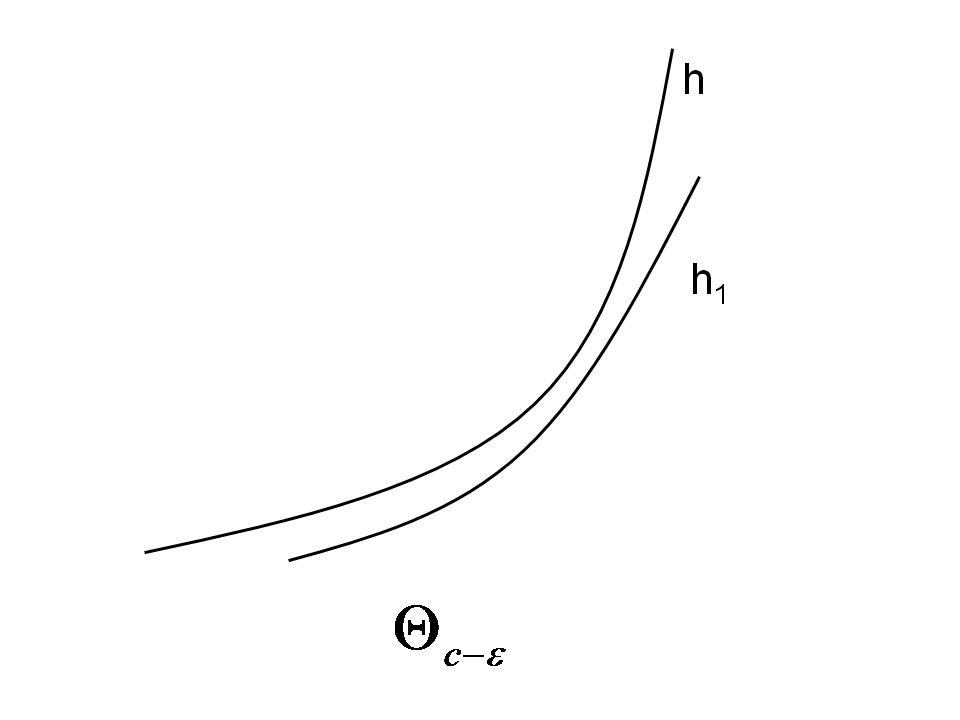}
    \includegraphics[width=50mm, bb=0 0 900 600]{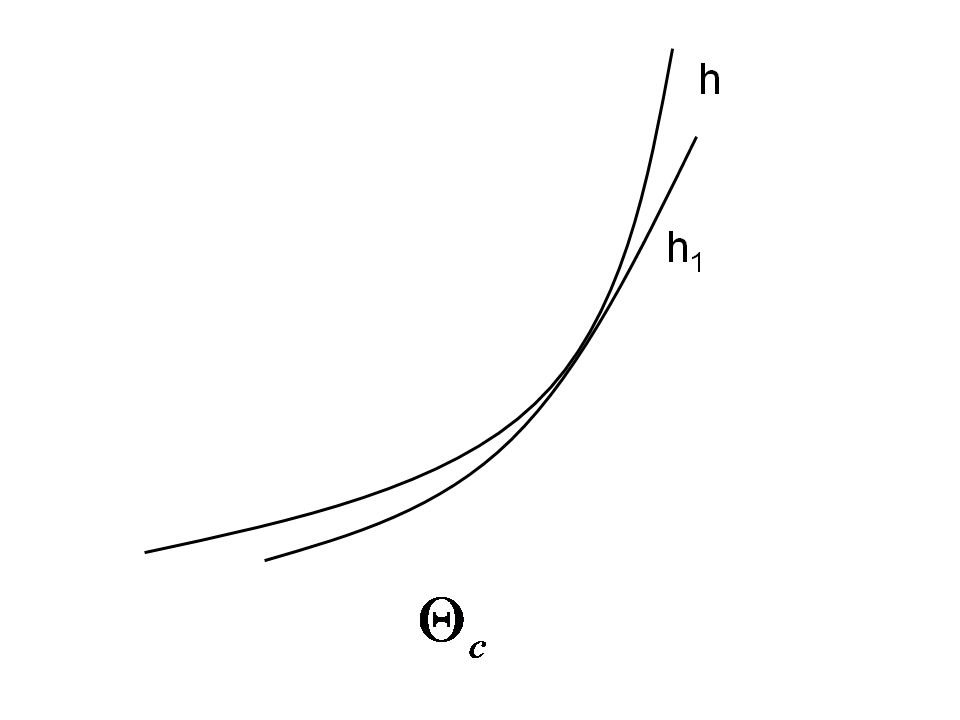}
    \includegraphics[width=50mm, bb=0 0 900 600]{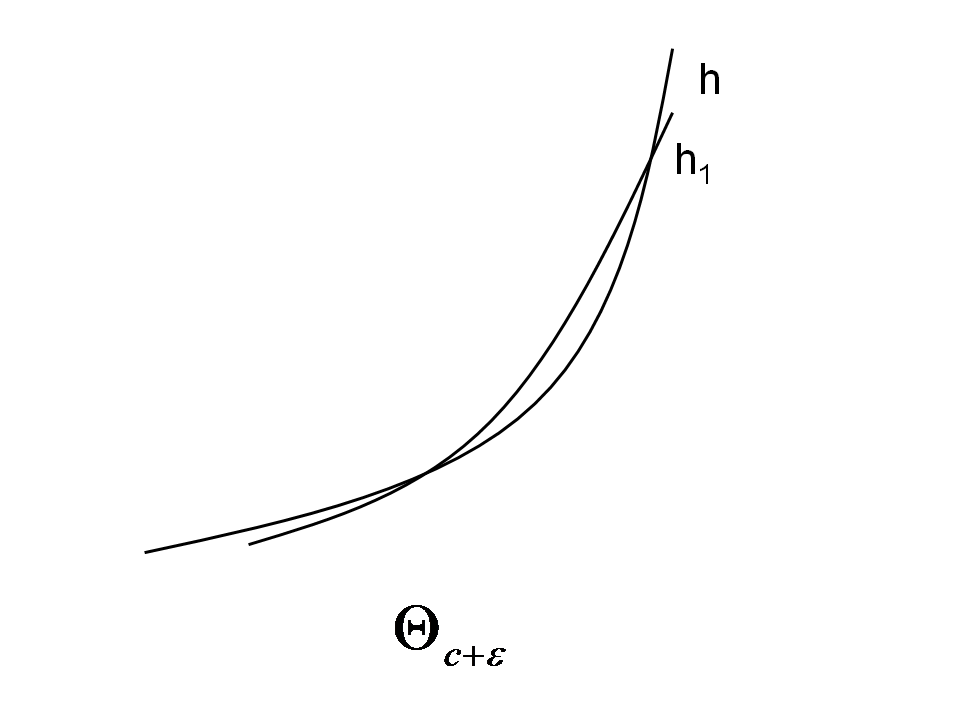}
\end{center}
\caption{One hyperbolic arc splits another (Case A1). The hyperbolic
arc $h_1$, the only element of $\I$, is closer than $h$ to the
$x$-axis on which an orthographic viewer is ``walking". Hence, $h_1$
moves faster and is visible (and enters the upper envelope) after
$x=c$.} \label{fig:CaseA1}
\end{figure}

\begin{figure}
\begin{center}
    \includegraphics[width=50mm, bb=0 0 900 600]{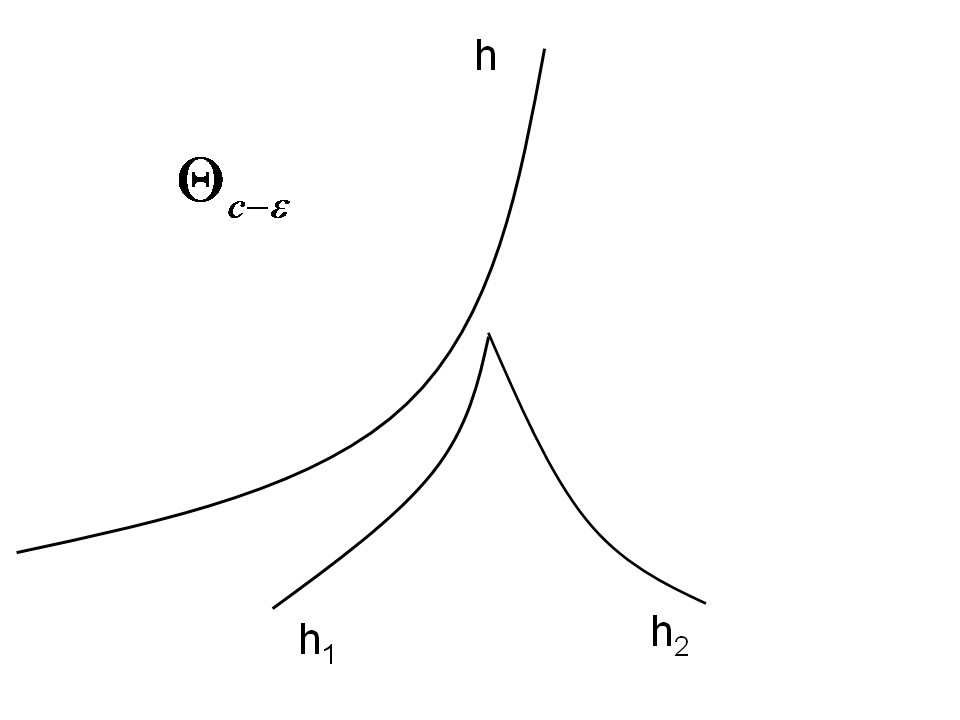}
    \includegraphics[width=50mm, bb=0 0 900 600]{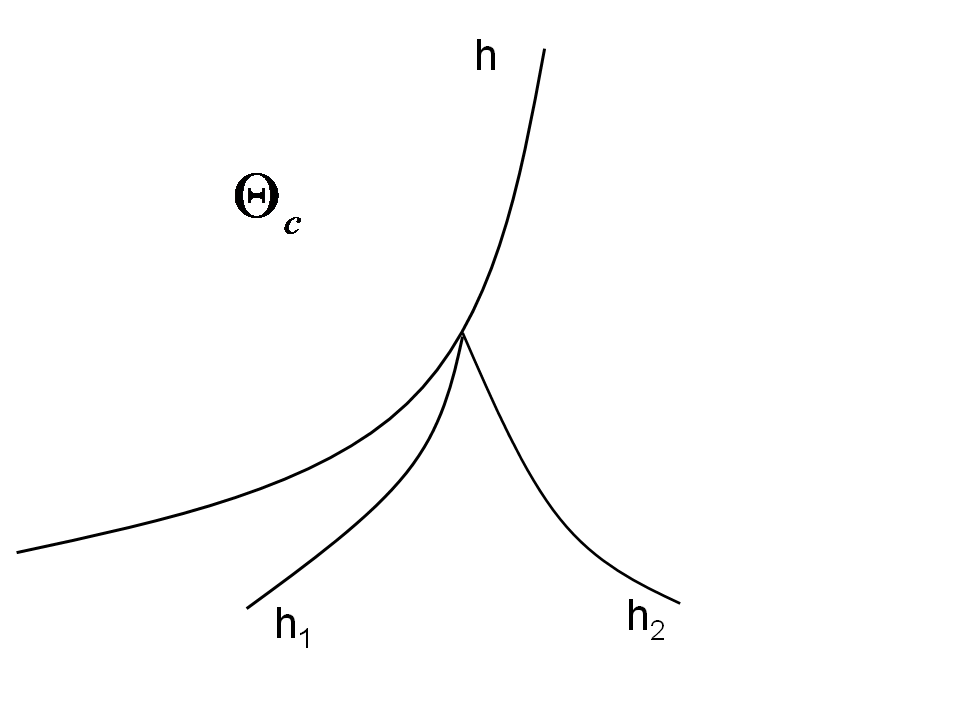}
    \includegraphics[width=50mm, bb=0 0 900 600]{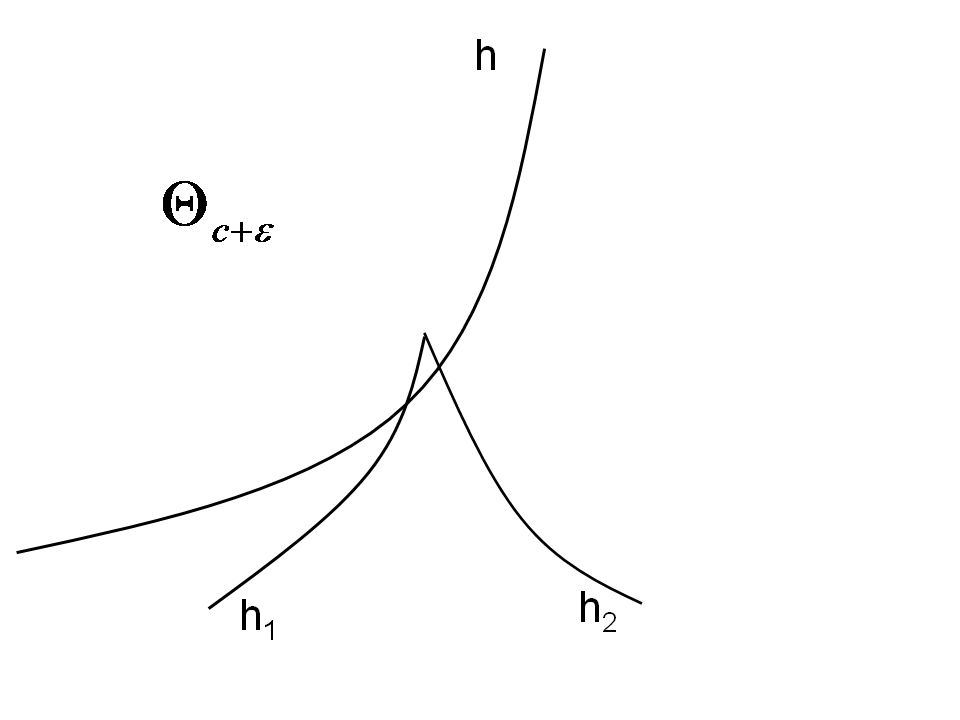}
\end{center}
\caption{Two hyperbolic arcs inserted in the middle of another
hyperbolic arc (Case A2). The hyperbolic arcs $h_1$ and $h_2$, which
make up the set $\I$, are closer  than $h$ to the $x$-axis on which
an orthographic viewer is ``walking". Hence, they moves faster and
are visible (and enter the upper envelope) after $x=c$.}
\label{fig:CaseA2}
\end{figure}

\begin{figure}
\begin{center}
    \includegraphics[width=50mm, bb=0 0 900 600]{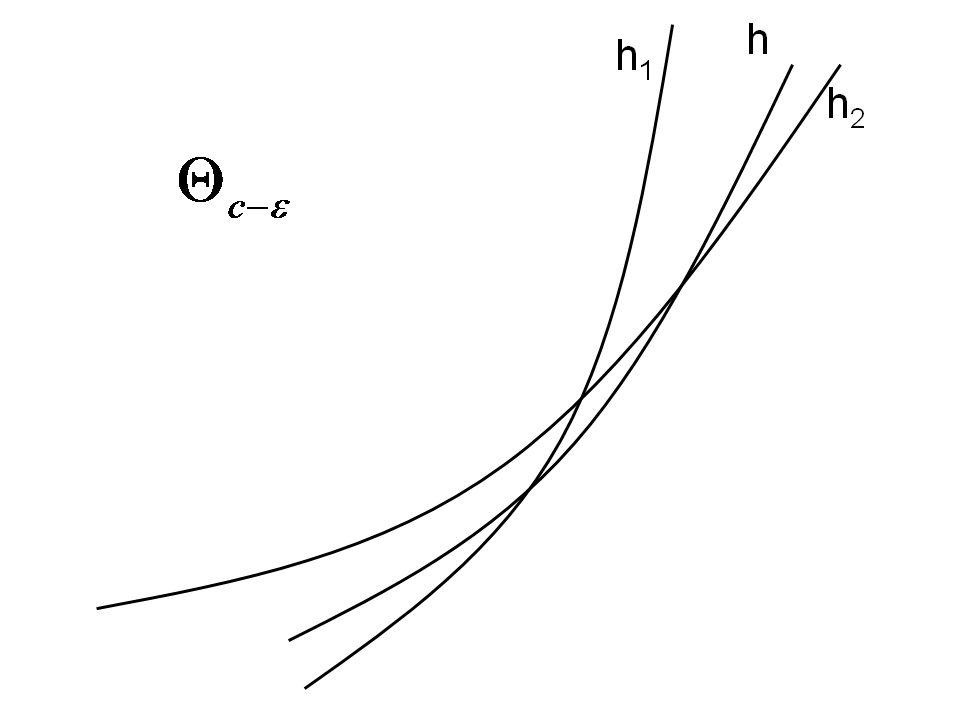}
    \includegraphics[width=50mm, bb=0 0 900 600]{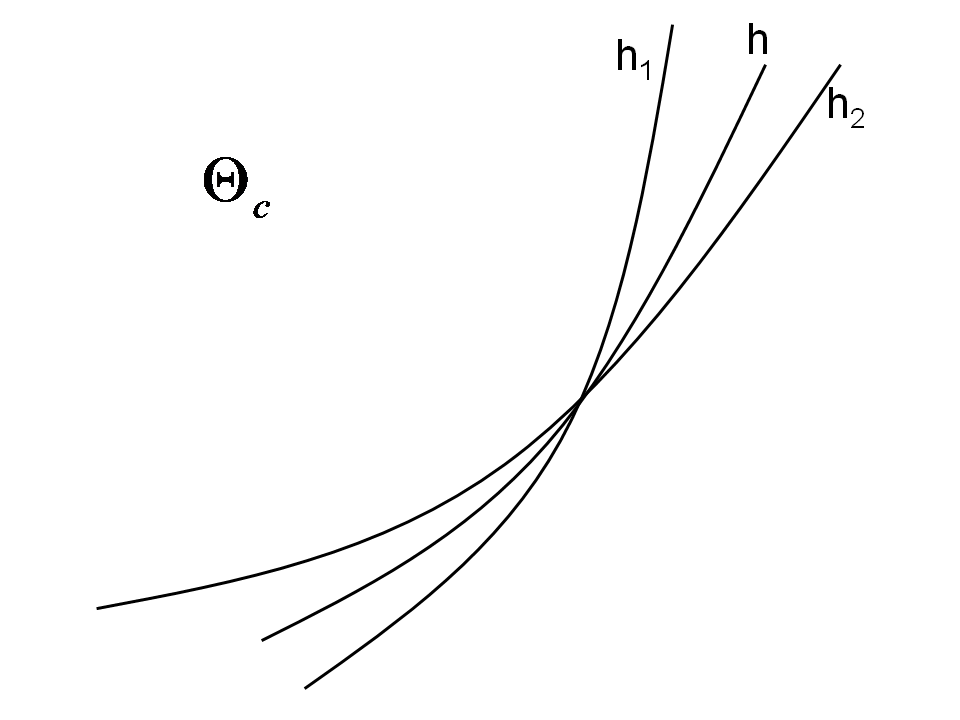}
    \includegraphics[width=50mm, bb=0 0 900 600]{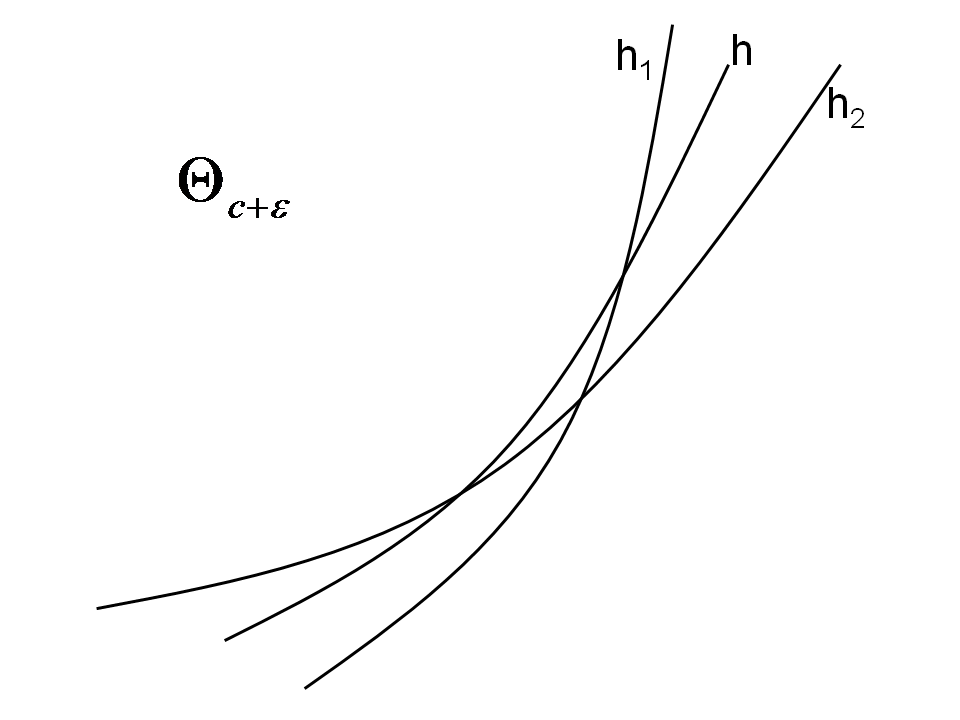}
\end{center}
\caption{A single hyperbolic arc is inserted between two other
hyperbolic arcs (Case B1). The hyperbolic arc $h$, the only element
in $\I$, is closer than $h_1$ and $h_2$ to the $x$-axis on which an
orthographic viewer is ``walking". Hence, $h$ moves faster and is
visible (and enters the upper envelope) after $x=c$.}
\label{fig:CaseB1}
\end{figure}

\noindent {\bf Case A*:} In this case, at our event point $x=c$,
three or
  more hyperbolic arcs that were occluded by the hyperbolic arc $h$ (and hence not present
  in $\L_{c-\e}$) appear in $\L_{c}$. This
  can happen only if the query line (passing through $x=c$ in 2D) at the appropriate angle,
  when
dropped from $z=+\infty$, will land on the three (or more) newly
appearing hyperbolic arcs and the occluding hyperbolic arc $h$,
thereby leading to four or more LECs, the degenerate case.

\noindent {\bf Case B*:} If this case were to occur, two hyperbolic
arcs $h_1'$ and $h_2'$ must be
  inserted between two other hyperbolic arcs $h_1$ and $h_2$ that were present in
  $\L_{c-\e}$. At such an event point $x=c$, the query line that
  intersects
  the $x$-axis at $x=c$ and lands on $h_1$ and $h_2$
  will also land on the other two hyperbolic arcs $h_1'$ and $h_2'$, thereby leading
  to four LECs.
  \qed \end{pf*}

We now turn our attention to the second and third questions: how
many event points can we encounter on the $x$-axis, and how do we
compute them?  A natural consequence of Lemma
\ref{lem:possibilities} is that each event point is defined by
either two or three hyperbolic arcs. Lemma \ref{lem:smallcases} allows
us to compute event points when we consider just two or three edges
in a Voronoi diagram. Therefore, in order to construct all the event
points, we have to consider all the hyperbolic arc subsets of size two
and three.

\begin{lem}
\label{lem:smallcases} All event points induced by a subset of
hyperbolic arcs containing either two or three arcs can be computed
in $O(1)$ time.
\end{lem}

\begin{pf*}{Proof sketch.} For the purpose of
this proof sketch, we rely on the orthographic viewing of the
hyperbolic arcs. Figure \ref{fig:CaseA1} illustrates the interaction
of two hyperbolic arcs to induce a change in the landing sequence.
With some algebraic and geometric manipulations, we find the point
$x=c$ where the hyperbola $h_2$ just appears. In similar fashion, we
can find the event point at which the landing sequence changes for
cases A2 and B1 also (see Figures \ref{fig:CaseA2} and
\ref{fig:CaseB1}). \qed \end{pf*}

\begin{lem}
\label{lem:seqChangeCard}  The number of the events on the $x$-axis
from the origin to $(0,1)$ is $O(n^3)$ and the ordered sequence of
events can be computed in $O(n^3 \log n)$ time.
\end{lem}

\begin{pf*}{Proof.} Cases A1,
A2, and B1 are all induced by the interaction of either two or three
hyperbolic arcs. Therefore, we consider all possible pairs and
triples and compute  the set of all the events that each pair or
triple can induce between $[0,1]$. The union of all these sets will
give us the set of all events. Since there are $O(n^3)$ triples and
fewer pairs and given Lemma \ref{lem:smallcases}, we can compute the
set of all events in $O(n^3)$ time. Therefore, the sorted {\em
sequence} of events can be computed in $O(n^3 \log n)$ time. \qed
\end{pf*}

In Lemma \ref{lem:seqChangeCard}, we show that the $x$-axis can be
divided into $O(n^3)$ intervals such that each is an instance of the
``query line through a point" case. This easily leads to a naive
$O(n^4\alpha(n)^{O(\alpha(n))} \log n)$ time algorithm in which we
store the $O(n^3)$ difference landing sequences, each of length at
most $O(n \alpha(n)^{O(\alpha(n))}\log n)$. In the query phase, we
can find the appropriate landing sequence in $O(\log n)$ time and
subsequently search within the landing sequence using the technique
described in Section \ref{sec:QuerySequence}, again, in $O(\log n)$
time.

We can do better using persistent data structures \cite{DSST89}. We
start with a landing sequence at the origin. The sequence takes at
most $O(n \alpha(n)^{O(\alpha(n))}\log n)$ space for the origin. It
gets updated at most at each of the $O(n^3)$ event points taken in
sorted sequence. When we calculate each event, we also store the
update that takes place; each update requires $O(1)$ space. In
particular, we store the following information:
\begin{enumerate}
    \item the $x$ value at which the event occurs,
    \item the sweep angle at which the event occurs,
  \item insertion or deletion,
  \item the sequence of hyperbolic arcs that are inserted or
  deleted, and
  \item the meeting numbers to ensure that the landing sequences can
  be searched in $O(\log n)$ time.
  \end{enumerate}
For the same reason that the size of the landing sequence is $O(n
\alpha(n)^{O(\alpha(n))}\log n)$ at the origin, the size never gets
beyond that at subsequent events. Therefore, with the above
information, we can use persistent data structures~\cite{DSST89} to
store the information in $O(n^3 +n \alpha(n)^{O(\alpha(n))}\log n) =
O(n^3)$ space. For each update, we have to perform an $O(\log n)$
search to find the appropriate location to make the update. There
can also be a constant number of updates in the meeting numbers in
the sequence. Each could require a binary search, taking $O(\log n)$
time. Therefore, the running time for the preprocessing phase is
$O(n^3 \log n)$.

\begin{algorithm}[h]
\caption{Preprocessing $P$ for arbitrary query lines.}
\floatname{algorithm}{Procedure} \label{alg:PreArb}
\begin{algorithmic}[1]

\STATE Construct the set $E$ of edges from the Convex Hull and the
Voronoi diagram for $P$.

\STATE Construct the hyperbolic arcs $H$, each corresponding to an
edge in $E$.

\STATE Compute  the sorted sequence of event points by considering
triples and pairs of hyperbolic arcs; additionally store the update
information along with each event point.

\STATE Compute $\L_0$, the landing sequence at the origin and store
it in a balanced linked binary search tree structure~\cite{Knuth73}.
Embed this linked structure into a persistent data structure
\cite{DSST89}.

\STATE For each event point taken in sequence, update the persistent
data structure according to the update information.
\end{algorithmic}
\end{algorithm}

\begin{algorithm}[h]
\caption{Querying with arbitrary query lines.}
\floatname{algorithm}{Procedure} \label{alg:QueryArb}
\begin{algorithmic}[1]

\STATE Find the $x$-intercept, $q_x$, of the query line.

\STATE Find the event point $c$ to the left of the $x$-intercept
such that $(c,q_x]$ is devoid of event points.

\STATE Find the binary search tree version corresponding to event
point $c$ and perform a binary search for the angle $\theta$ that
the query line makes with the $x$-axis.
\end{algorithmic}
\end{algorithm}

In the query phase, we search for the event point that just preceded
the $x$ intercept of the query line; this takes $O(\log n)$ time. We
lookup the landing sequence induced by the event point, which again
takes $O(\log n)$ time. This will also be the landing sequence at
the $x$ intercept. Armed with the landing sequence, we binary search
in $O(\log n)$ time to find the exact hyperbolic arc on which the
query line lands. This leads us to our final result:

\begin{thm}
\label{thm:final} Given a set $P$ of $n$ points in $[0,1]^2$, we can
preprocess it in time $O(n^3 \log n)$ and space $O(n^3)$ such that
when presented with an arbitrary query line $Q$, we can report the
LEC centered on $Q$ and within the convex hull of $P$ in time
$O(\log n)$.
\end{thm}

We stated earlier that we consider the case where more than three
LECs occur on a query line to be degenerate. This is an essential
requirement for Lemma \ref{lem:possibilities}. It can be detected
when we consider pairs and triples of hyperbolic arcs to compute the
event points. If more than one event occurs at the same point, say
$x=c$ on the $x$-axis, and at the same angle, we can conclude that
this degeneracy has occurred. We can avoid this degeneracy by
slightly perturbing the points in $P$ that induce the four or more
LECs that cause the degeneracy.

\noindent {\bf Acknowledgements: } We are indebted to M. V.
Panduranga Rao and Leo Livshits for discussing useful ideas. We are
also grateful to the anonymous reviewers of an earlier manuscript
that helped make the current version more readable.

\bibliographystyle{abbrv}
\bibliography{papers}

\end{document}